\documentstyle{article}
\title{\bf Time Walk Through the Quantum\\
Cosmic String}
\author{ Pedro F. Gonz\'alez-D\'{\i}az.\\
Instituto de Matem\'aticas y F\'{\i}sica Fundamental\\
Consejo Superior de Investigaciones Cientificas\\
Serrano 121, 28006 Madrid (SPAIN)\\
}
\date{December 10, 1994}
\begin{document}
\maketitle
\large
\setlength{\baselineskip}{0.5cm}

\begin{center}
{\bf Abstract}
\end{center}

\noindent This paper deals with the geometry of supermassive cosmic
strings and its possible connection with the existence of closed
timelike curves (CTCs), so as with inflationary expansion. The linear
energy density, $G\mu$, of supermassive strings becomes so large
that gravitational effects dominate,
and therefore we have used an approach that enforces the
spacetime of these strings to also satisfy the symmetry of a
cylindric gravitational topological defect, that is a spacetime
kink. In the simplest case of kink number unity the entire
energy range of supermassive strings becomes then quantized so
that only cylindrical defects with $G\mu=\frac{1}{4}$
(critical string) and $G\mu=\frac{1}{2}$ (extreme
string) are allowed to occur
in this range. It has been seen that the internal spherical
coordinate $\theta$ of the string also evolves on imaginary
values, leading to the creation of a covering shell of broken phase
that protect the core with trapped energy, even for $G\mu=
\frac{1}{2}$. Then, the conical singularity becomes a
removable horizon singularity. We re-express the extreme
string metric in the Finkelstein-McCollum standard form and
remove the geodesic incompleteness by using the Kruskal
technique. A $z$=const section of the resulting metric
is the same as the $\theta=\frac{\pi}{2}$ section of a
De Sitter kink, though it requires additional coordinate
regions to completely describe it. It is shown that through
such new regions the extreme string spacetime can accommodate
CTCs. We discuss two of such curves and conclude that quantum
effects allow them to exist, though it is also seen by using
path integral formalism that quantum theory prevents these
CTCs to be noticed by any observer. Also discussed is the
idea that the extreme string is the unique local string
which is able to spontaneously drive
inflationary expansion in its core,
without any fine tuning of the initial conditions.

\pagebreak

\section{\bf Introduction}

Topological defects consisting of confined regions of false vacuum,
trapped inside bubbles of true vacuum, can occur in gauge theories
with spontaneous symmetry breaking [1]. Among such defects, cosmic
strings arising in the breaking of a local $U(1)$ gauge symmetry
are of particular interest [1,2]. If local strings appeared in
phase transitions taking place in the early universe, they could
have served as seeds for the formation of galaxies and other
larger-scale structures we now are able to observe [3]. Recently,
the possibility that inflation can be driven in the core of
topological defects has also been advanced [4,5]. In the case of
local cosmic strings this extremely interesting possibility will
critically depend [5] on the nature and strength of their gravity
coupling $G\mu$, where $\mu$ is the string mass per unit length.
It appears that only when the symmetry-breaking scale
approaches the extreme supermassive limit $G\mu=\frac{1}{2}$, the
size of the false vacuum region inside the string may exceed the
horizon size that corresponds to the vacuum energy in the core
and drives an inflationary process.

This can be seen for static defect solutions in simple models
\[V(\varphi)=\frac{1}{4}\lambda(\varphi_{a}\varphi_{a}-\eta^{2})^{2},
a=1,...,N.\]
Cosmic strings ($N=2$) cannot exist for $G\mu\geq 1$ [6].
However, this fact is by no means meaning that inflation cannot be
driven by local strings. Equalizing the core radius $\delta_{0}
=\eta (\pi V_{0})^{-\frac{1}{2}}$,
where $V_{0}=\frac{1}{4}\lambda\eta^{4}$,
and the horizon size corresponding
to the vacuum energy $V_{0}$,
$H_{0}^{-1}=(\frac{3}{8\pi GV_{0}})^{\frac{1}{2}}$,
one obtains using $\mu=\eta^{2}$ a
value for the string mass per unit length $\mu_{I}=
\frac{3}{8G}$, such that for $\mu>\mu_{I}$ inflation can be
generated in the core.

Nevertheless, the concepts of string radius and
mass per unit length for a source
like the string core are not unambiguously defined. The
above expression for $\delta_{0}$, which corresponds to
the radius of the string cylinder, and the usual
definition [7] of $\mu$,
by which one simply integrates the uniform
energy density of the string interior over the proper volume of
the core, become only
unambiguous for a pefectly spherical string interior
i.e. when $2G\mu=2G\eta^{2}=1$. It would then be expected that
as one separates from a perfect spherical string for
$G\mu<\frac{1}{2}$, the effective string radius would
become somewhat smaller than that is predicted by
$\delta_{0}=\eta(\pi V_{0})^{-\frac{1}{2}}$, so that
inflation could be induced in the string core
only  when $\mu=\mu_{e}=\eta^{2}=\frac{1}{2G}$, or for values
of $\mu$ very close to $\mu_{e}$.
Unfortunalety, an extreme
supermassive string with $2G\mu_{e}=1$ does not seem to
exist because it would correspond to a situation where all
the exterior broken phase is collapsed into the core, leaving
a pure false-vacuum phase in which the picture of a cosmic
string with a core region of trapped energy is lost [6-8].
This happens in all considered string metrics, i.e. for the
Hiscock-Gott metric [10,11] and for the Laguna-Garfinkle [8]
and Ortiz [9] metrics, in all the cases the spacetime possessing
an unwanted singularity which cannot be smoothed out.

Strings for which $4G\mu<1$ distort the underlying spacetime
structure, but their own internal structure
is not essentially modified
by the structure of the underlying spacetime [7].
For such strings, spacetime is somehow adapted to
the topological defect imposed by the matter gauge fields,
resulting in a conical geometry, but this does not backreact
over the string structure.
This is no longer the case, however, for supermassive strings [8,9],
$4G\mu\geq 1$, where the strength of gravity coupling is so
great and its nature such that one should expect the string
structure itself to be substantially
modified by the characteristics it induces in the
underlaying spacetime as well; that is to say, one should expect
supermassive cosmic strings to be those stable topological
defects which smoothly follow the structural pattern dictated
by a gravitational kink, i.e. the pattern of an allowed
gravitational topological defect which can move about
spacetime but cannot be removed without cutting [12].

By enforcing the interior metric of the string to follow the
spacetime lightcone itinerary dictated by a cylindrically-symmetric
gravitational kink, we show in this paper that the picture of a
cosmic string with a core region of trapped energy is still
retained even at the extreme value $2G\mu=1$, and that the
unwanted conical singularity becomes the apparent singularity
(event horizon) of a De Sitter kink. This horizon singularity
can now be removed by a suitable Kruskal extension. The
resulting extreme supermassive string is then able to drive
an essentially unique, gravitational inflationary process,
without any fine tuning of the initial conditions. Regarding
the extreme supermassive strings as initial baby universes,
this inflationary expansion of the string core can then lead to
a {\it unique} stationary picture of eternally self-reproducing
inflating universes [13]. It is also shown that the extreme
supermassive string model can accommodate
finite spacetime closed timelike
curves [14], without any imaginary mass states or causality
violation [15].

The paper is outlined as follows. In section 1 we construct
an explicit metric for the supermassive cosmic string kink,
and discuss the constraints that the existence of the kink
imposes on the internal geometry of the string. As a result
of the surface identifications arising from the constraints,
the energy of the supermassive cosmic string becomes quantized
so that it can only take on values $4G\mu=1$ and
$2G\mu=1$. We then re-express the metric of the cosmic
string kink in standard form and obtain an analytical expression
for the relevant time parameter entering that metric in section
3, where a discussion on surface identifications in kink space
is moreover included. The geodesic incompleteness of the
standard metric is removed in section 4 by maximally-extending
this metric using the Kruskal technique. In section 4 we also
show that the geodesically complete metric of a $2G\mu=1$
cosmic-string kink describes a given section of a kinky De
Sitter spacetime which can only be represented by using four
different coordinate regions, while allowing the existence
of an eternal inflationary process. Finally, we discuss the
possibility for the existence of closed timelike curves in
the above Kruskal spacetime first, in section 5, by
explicitely determining the null-geodesic itinerary followed
by two of such curves, and then, in section 6, by studying
the quantum creation of particles induced by propagators
along the null geodesics. It is concluded that although closed
timelike curves are allowed to exist, no observer could ever
notice their existence. Throughout the paper we use units
so that $\hbar=c=1$.

\section{\bf The Cosmic String Kink}
\setcounter{equation}{0}

The static, cylindrically symmetric internal metric of a straight
cosmic string is [10,11]
\begin{equation}
ds^{2}=-d\tau^{2}+d\rho^{2}+dz^{2}+r_{*}^{2}\sin\frac{\rho}{r_{*}}d\phi^{2},
\end{equation}
with $-\infty<\tau<\infty$, $-\infty<z<\infty$, $0\leq\phi<2\pi$,
$0\leq\rho\leq r_{*}\arccos(1-4G\mu)$, and
\begin{equation}
r_{*}=(8\pi G\epsilon)^{-\frac{1}{2}},
\end{equation}
where $\epsilon$ is the uniform string density, out to some
cylindrical radius $\rho_{0}$.

Both the interior metric (2.1) and the exterior metric,
\begin{equation}
ds^{2}=-d\tau^{2}+d\rho'^{2}+dz^{2}+(1-4G\mu)^{2}\rho'^{2}d\phi^{2},
\end{equation}
define two surfaces at $z=$const, $\tau=$const, which can be
simultaneously visualized by embedding the metrics in an Euclidean
three-space [11]. Then, the geometries of such surfaces are,
respectively, that of a spherical cap (interior region) and that
of a cone with deficit angle $\Delta=8\pi G\mu$ in the exterior
vacuum region.

The change of coordinate
\begin{equation}
r=r_{*}\sin\frac{\rho}{r_{*}}
\end{equation}
shows more transparently the invariance of the line element (2.1)
under rotation about $\phi$ and traslation along $z$ of the
underlying cylindrical symmetry (i.e. under the two Killing
vectors $\xi=\partial_{\phi}$ and $\zeta=\partial_{z}$ [16]),
but gives rise to a singularity at $r=r_{*}$; that is
\begin{equation}
ds^{2}=-d\tau^{2}+\frac{dr^{2}}{1-\frac{r^{2}}{r_{*}^{2}}}+dz^{2}+r^{2}d\phi^{2}.
\end{equation}

Clearly, the divergence of (2.5) at $r=r_{*}$ must correspond to
an apparent singularity and should therefore be removable by an
appropriate coordinate transformation. A coordinate change which
still transparently shows invariance under the Killing vectors
and leads to no interior singularity is
\begin{equation}
u=\frac{\tau}{r_{*}}+\arcsin\frac{r}{r_{*}},
v=\frac{\tau}{r_{*}}-\arcsin\frac{r}{r_{*}}.
\end{equation}
In terms of coordinates $u$ and $v$ we obtain for the interior metric
\begin{equation}
ds^{2}=-r_{*}^{2}dudv+dz^{2}+r^{2}d\phi^{2},
\end{equation}
where
\begin{equation}
r=r_{*}\sin[\frac{1}{2}(u-v)].
\end{equation}
In any case, the exterior metric still keeps a conical singularity.

The regular metric (2.7) would trivially describe the maximally
extended interior region of a cosmic string when $4G\mu<1$.
The structure of the string core is then determined by
the matter gauge theory (e.g the Abelian Higgs model of Nielsen
and Olesen [17]) rather than gravity, and should only lodge the
real energy from the unbroken gauge phase. Nevertheless,
according to that was discussed in the Introduction, at scales
where the gravity coupling is very large, $4G\mu\geq 1$
[8-11], the cosmic string should also satisfy the conservation
law for topological defects which are described by relativistic
metric twists, and hence the coordinate $r$ entering metric (2.5)
must vary along an extended interval in such a way that it
allows metric (2.5) to also become a cylindrically-symmetric
spacetime kink. In what follows we shall see that, at such
scales, the maximally extended internal metric will also cover
a given strip with the broken phase, surrounding the core of
trapped real energy. These two internal phases are then
separated by an event horizon at $r=r_{*}$.

The relationship between a possible cosmic string kink and
the apparent singularity in metric (2.5) can be appreciated
by considering [18] a cylindrically-symmetric spacetime with
manifest invariance under the action of the Killing vectors
$\xi$ and $\zeta$ of the form
\begin{equation}
ds_{\alpha}^{2}=\cos 2\alpha(-dt^{2}+dr^{2})-2\sin 2\alpha
dtdr+dz^{2}+r^{2}d\phi^{2},
\end{equation}
where $\alpha$ denotes the angle of tilt of the spacetime
lightcones. We shall then ensure the presence of one kink
(i.e. a gravitational topological defect with kink number
unity) by requiring $\alpha$ to monotonously vary between
$0$ and $\pi$, in such a way that $\alpha$ will either
increase from $0$ to $\pi$ if this is the direction along
which the time entering the relevant (explicitely showing
the kink) metric increases, or decreases from $\pi$ to $0$,
otherwise.

We note that (2.9) can be transformed into (2.5) if we set
\begin{equation}
\sin\alpha=\frac{r}{2^{\frac{1}{2}}r_{*}},
\end{equation}
and change the time coordinate so that $\tau=t+G(\omega)$, where
\begin{equation}
d\omega=d\tau+F(r)dr,
\end{equation}
with $F(r)$ a given function of $r$. Denoting $\frac{dG(\omega)}{d\omega}$
by $G'$, it follows
\begin{equation}
dt=d\tau(1-G')-G'F(r)dr.
\end{equation}
Metric (2.5) can then be obtained from (2.9) if
\begin{equation}
G'=1-\frac{k_{1}}{\cos^{\frac{1}{2}}2\alpha}, k_{1}=\pm 1
\end{equation}
\begin{equation}
G'F(r)=\tan 2\alpha.
\end{equation}
Note that both $F(r)$ and $G'$ are singular at $r=r_{*}$. This simply
reflects the singular character of metric (2.5).

Before discussing the above transformation, let us consider the
limitations that the functional form (2.10) imposes on the very
internal geometry of a supermassive cosmic string. From (2.4) and
(2.10) it is obtained
\begin{equation}
\cos^{2}\theta = \cos 2\alpha,
\end{equation}
where we have denoted $\theta=\frac{\rho}{r_{*}}$, and $\alpha$
monotonously varies between $0$ and $\pi$.

The question now is,
how does $\theta$ vary under variation of $\alpha$ between $0$
and $\pi$?. From (2.15) it follows that monotonous variation of
$\alpha$ from $0$ to $\frac{\pi}{4}$, and from $\frac{3\pi}{4}$
to $\pi$ induces monotonous variation of $\theta$, respectively,
from $0$ to $\frac{\pi}{2}$ and from $\frac{\pi}{2}$ to $\pi$
(or, likewise, from $\pi$ to $\frac{\pi}{2}$ and from $\frac{\pi}{2}$
to $0$). Note that induced variations of $\theta$ in the interval
($\pi$,$2\pi$) are also allowed but, since $\theta$ appears in the
metric in the form $\sin^{2}\theta$, these variations would lead to
the same geometrical situations as from variations of $\theta$
in the interval ($0$,$\pi$). We obtain then that if $\alpha$ is
allowed to run along its entire interval from $0$ to $\pi$, and
since the kink metric (2.9) depends on $\alpha$ only through the
argument $2\alpha$ of the $\cos^{2}$ and $\sin$, the two opposite
poles of the resulting interior two-sphere at $\theta=0$ and
$\theta=\pi$ should be identified.

Variation of $\alpha$ from $\frac{\pi}{4}$ to $\frac{3\pi}{4}$
induces variation of $\theta$ only along its imaginary axis,
first from $\frac{\pi}{2}$ to $\frac{\pi}{2}+i\ln(2^{\frac{1}{2}}\pm 1)$
(at $\alpha=\frac{1}{2}$), and then to $\frac{\pi}{2}$ again.
The choice of sign in the argument of the $\ln$ at the extremum
value of $\theta$ corresponding to $\alpha=\frac{\pi}{2}$ should
be made as follows. On the interval where $\theta$ is complex,
$\theta=\frac{\pi}{2}+i\theta_{i}$, we have (see Refs. [10,11])
\begin{equation}
G\mu=\frac{1}{4}+\frac{i}{4}\sinh\theta_{i}
\end{equation}
\begin{equation}
r=ir_{*}\coth\theta_{i}
\end{equation}
Taking into account (2.17) it follows that the imaginary part of
(2.16) corresponds to a real mass per unit imaginary length, or
equivalently, a tachyonic mass per real unit length. Thus, once
the critical mass $G\mu=\frac{1}{4}$ is reached, the interior of
the string starts developing a new real region with maximum width
$(2^{\frac{1}{2}}-1)r_{*}$, covering the entire hemisphere, where
the gauge symmetry is broken at the scale $\varphi=\pm\eta$. Then,
the symmetry $\eta\rightarrow-\eta$ of the broken phase will
imply an identification $\frac{\pi}{2}+i\ln(2^{\frac{1}{2}}+1)$
$\rightarrow\frac{\pi}{2}+i\ln(2^{\frac{1}{2}}-1)$ on the extremum
value of $\theta_{i}$ that corresponds to $\alpha=\frac{\pi}{2}$.
We can therefore choose complex $\theta$ to vary first from
$\frac{\pi}{2}$ to $\frac{\pi}{2}+i\kappa\ln(2^{\frac{1}{2}}+1)$
as $\alpha$ goes from $\frac{\pi}{4}$ to $\frac{\pi}{2}$, and
then from $\frac{\pi}{2}+i\kappa\ln(2^{\frac{1}{2}}-1)$ to
$\frac{\pi}{2}$ again as $\alpha$ goes from $\frac{\pi}{2}$ to
$\frac{3\pi}{4}$, either for $\kappa=\pm 1$ (Fig. 1).

Now, the maximum value of the internal spherical coordinate
$\theta$, $\theta_{M}$, is related to the string mass per
unit length $\mu$ by [10,11]
\[\theta_{M}=\arccos(1-4G\mu).\]
Then, in order to ensure the occurrence of one kink in the
cosmic string interior, we must have $\theta_{M}=\pi$ and hence
$G\mu=\frac{1}{2}$; or in other words, the gravitational
coupling implies a {\it quantization} of supermassive cosmic
strings so that only the critical hemispherical string at
$G\mu=\frac{1}{4}$ (no kink present), and the extreme
spherical string at $G\mu=\frac{1}{2}$ (one kink present)
are allowed to exist along the entire interval,
$\frac{\pi}{2}\leq\theta\leq\pi$, of possible classical
supermassive cosmic strings. We regard this as a typical
quantum-gravity effect.

If we ensure the presence of a kink, then no string exterior
(except the portion which is incorporated into the interior
metric by the complex evolution of $\theta$) is possible or
needed [10,11], and the conical singularity becomes a removable
horizon singularity on the core surface at $r=r_{*}$,
separating the two gauge phases that make up the extended
string interior. Thus, all of the possible geometry of
the extreme string can be regarded as describable by a
spacetime which is ${\bf R}^{2}\times S^{2}$, showing still
the picture of a cosmic string with a spherical core region
of trapped energy surrounded by a shell of true vacuum
protecting the string from dissolving in the unbroken
symmetry phase.

\section{\bf Lightcone Configurations}
\setcounter{equation}{0}

The results of section 2 allow us to deal with the interior metric
of an extreme supermassive cosmic string (hereafter denoted as
"extreme string") in a similar fashion to as it is done in the cases
of the black hole kink [18] or the De Sitter kink [19].

The interior of an extreme string has still a geodesic incompleteness
at $r=r_{*}$, and can only be described by using a number of
different coordinate patches. The identification  and
distinction of such patches can be achieved by transforming
(2.9) into the standard metrical form proposed by
Finkelstein and McCollum [18], adapted to cylindric symmetry.
To accomplish such a transformation, it is convenient to
introduce a new time coordinate, such that
\begin{equation}
\bar{t}=t+G(\sigma)
\end{equation}
\begin{equation}
\cos 2\alpha(1-G^{;2}F_{\sigma}(r))+2\sin 2\alpha G^{;}F_{\sigma}(r)=0,
\end{equation}
where the new variable $\sigma$ and the new functionals $F_{\sigma}(r)$
and $G\equiv G(\sigma)$ are defined as follows
\[d\sigma=d\bar{t}+F_{\sigma}(r)dr\]
\[G^{;}\equiv G(\sigma)^{;}=\frac{dG}{d\sigma}.\]
We obtain then
\begin{equation}
G^{;}F_{\sigma}(r)=\tan 2\alpha-\frac{k_{2}}{\cos 2\alpha}, k_{2}=\pm 1,
\end{equation}
so that metric (2.9) becomes
\begin{equation}
ds^{2}=-d\bar{t}^{2}-\frac{2k_{1}k_{2}}{\cos^{\frac{1}{2}}2\alpha}d\bar{t}dr+dz^{2}+r^{2}d\phi^{2},
\end{equation}
where $k_{1}$ is defined in (2.13). Metric (3.4) still is not the kink
metric in standard form. This is obtained by making the re-definition
\begin{equation}
k_{1}d\tilde{t}=k_{1}\frac{d\bar{t}}{\cos^{\frac{1}{2}}2\alpha}=dt+(\tan
2\alpha-\frac{k_{2}}{\cos 2\alpha})dr,
\end{equation}
so that we finally obtain
\begin{equation}
ds^{2}=-\cos 2\alpha
d\tilde{t}^{2}-2k_{1}k_{2}d\tilde{t}dr+dz^{2}+r^{2}d\phi^{2}.
\end{equation}
Since one can always write $k_{1}k_{2}=k=\pm 1$, (3.6) becomes formally
the same as the general line element in standard form given by
Finkelstein and McCollum [18], after exchanging spherical for
cylindric symmetry. On the other hand, using (2.10) it can be
readily seen that any $z$=const section of metric (3.6) is not
but the $\theta=\frac{\pi}{2}$ (hemispherical)
section of the standard De Sitter
kink metric [19] for a positive cosmological constant
$\Lambda=\frac{3}{r_{*}^{2}}$.

However, the facts that the sign definitions given by parameters
$k_{1}$ and $k_{2}$ enter the formalism at different levels (Ref.
Eqns (2.13) and (3.3)), and that the time parameter entering
the standard metric for the De Sitter kink is defined in a
similar fashion to as it is made in (3.1), rather than (3.5),
leads to a sharp distinction between the geometries of
De Sitter kink and extreme string kink. In order to clearerly
appreciate the nature of that distinction, let us calculate
now $\bar{t}$ for the case $t=0$ where we enforce the system
to lie on the $r$ axis. This involves calculating the
following two integrals
\[I_{1}=\int_{0/A}^{r}\frac{dr}{\cos^{\frac{1}{2}}2\alpha}\]
\[I_{2}=\int_{0/A}^{r}\frac{dr\sin 2\alpha}{\cos^{\frac{1}{2}}2\alpha},\]
where the lower limit $0/A$ refers to the choices $r=0$ and
$r=A\equiv 2^{\frac{1}{2}}r_{*}$, depending on whether the
case $k_{2}=+1$ or the case $k_{2}=-1$ is being considered [19].
Taking
\begin{equation}
\sin\alpha=\frac{r}{A},
\cos\alpha=k_{2}(1-\frac{r^{2}}{2r_{*}^{2}})^{\frac{1}{2}},
\cos^{\frac{1}{2}}2\alpha=k_{1}(1-\frac{r^{2}}{r_{*}^{2}})^{\frac{1}{2}},
\end{equation}
we obtain
\begin{equation}
I_{1}=-ir_{*}\ln[i\frac{r}{r_{*}}+k_{1}(1-\frac{r^{2}}{r_{*}^{2}})^{\frac{1}{2}}]\mid_{0/A}^{r}
\end{equation}
and
\[I_{2}=-\frac{k_{2}r_{*}^{2}}{A}\{k_{1}[(1-\frac{r^{2}}{2r_{*}^{2}})(1-\frac{r^{2}}{r_{*}^{2}})]^{\frac{1}{2}}\]
\begin{equation}
+\frac{r_{*}}{A}\ln[\frac{2k_{1}A}{r_{*}}(1-\frac{r^{2}}{r_{*}^{2}})^{\frac{1}{2}}+4(1-\frac{r^{2}}{2r_{*}^{2}})^{\frac{1}{2}}]\}\mid_{0/A}^{r}.
\end{equation}

{}From (3.8) and (3.9) it can be seen that it is not possible to obtain
a unique, compact expression for the time
parameter $\bar{t}$. This fact
distinguishes the extreme string kink from the De Sitter kink
for which there is a well defined analytical definition of $\bar{t}$ [19].
The geometrical reason for this is that time parameter $\bar{t}$
actually corresponds to the case where
a complete description of the geometry requires
just the two patches associated with $k_{2}=\pm 1$,
which is the case for the De Sitter kink but not for
the extreme string kink. Because of the presence of factor
$\cos^{-\frac{1}{2}}2\alpha$ in the expression for $G'$ in
(2.13), each of the two coordinate patches associated with
$k_{2}=\pm 1$ itself must unfold in the two different
sets of coordinate
regions that correspond to the new sign ambiguity $k_{1}=\pm 1$.
That unfolding can only be accounted for by the new time
parameter $\tilde{t}$ entering the standard metric (3.6) which,
according to the sign definition introduced for $k_{2}$ in (3.5),
will correspond to a physical time $t$ running either
forwards when $k_{1}=+1$, and backwards when $k_{1}=-1$.
One should then expect a unique, compact analytical
expression governing all coordinate domains only for time
$\tilde{t}$. By directly integrating (3.5) between the same
limits as in (3.8) and (3.9) we in fact finally obtain [20]
\[\tilde{t}\equiv\tilde{t}(k_{1},k_{2})=\int_{0/A}^{r}d\tilde{t}=k_{1}t-r_{*}k_{1}k_{2}\{\frac{A}{r_{*}}(1-\frac{r^{2}}{2r_{*}^{2}})^{\frac{1}{2}}\]
\begin{equation}
-\frac{1}{2}\ln[\frac{(A(1-\frac{r^{2}}{2r_{*}^{2}})^{\frac{1}{2}}+r_{*})(r_{*}-r)}{(A(1-\frac{r^{2}}{2r_{*}^{2}})^{\frac{1}{2}}-r_{*})(r_{*}+r)}]\}.
\end{equation}
Note that all sign ambiguity arising from the square root in the
argument of the $\ln$ has been omitted in Eqn. (3.10). Such an
ambiguity does not affect the discussion to follow as it
only manifests as an additive constant term which
leaves metric (3.6) unchanged. This ambiguity will be of decisive
importance however for the consideration of the thermal properties
of the extreme string which we shall deal with in section 6.

The fact that for $t=0$ both $k_{1}$ and $k_{2}$ enter (3.6) and
(3.10) at exactly the same footing, so that all distinction between
the coordinate domains that arise from different values of $k_{1}$
disappears to leave just two different coordinate patches
corresponding to $k=k_{1}k_{2}=\pm 1$, is just an artifact coming
from still keeping the removable geodesic incompleteness in
metric (3.6). It will be seen in the next section that even for
$t=0$ the sign parameters $k_{1}$ and $k_{2}$ actually enter
the coordinates defining the maximally-extended metric at different
footings. Therefore, although $\alpha$ may only vary within the
interval ($0$,$\pi$), there really are four distinguishable
sets of
coordinate regions that correspond to the distinct combinations
of $k_{2}=\pm 1$ with $k_{1}=\pm 1$. In the general case $t\neq 0$
the coordinate regions can even be distinguished without removing
the event singularity.

In the De Sitter space, where the geometry of a $t$=const, $r$=
const section is just a two-sphere, the spherical symmetry is
manifested in the presence of a kink by just identifying the
equators of the two hemispheres that, respectively, correspond
to the two kink coordinate patches, but the poles of these
hemispheres can never be identified [18,19]. In contrast, the
real geometry of a $t$=const, $z$=const section of extreme
string kink is that of two real hemispheres whose mutual
matching at their equators lies on the extremum imaginary
values of the spherical coordinate $\theta$,
$\ln(2^{\frac{1}{2}}+1)\rightarrow\ln(2^{\frac{1}{2}}-1)$. The
imaginary values of $\theta$ are mapped onto
real values of $\alpha$ in the kink, giving rise to an equator
identification at $r=A$, provided the resulting values of $\alpha$
are either always increasing from 0 to $\pi$ (i.e. when $k_{1}=+1$)
as one goes along the two coordinate regions being joined from
$r=0$ to $r=0$, or always decreasing from $\pi$ to 0 (i.e. when
$k_{1}=-1$) along the two entire joined coordinate patches. Thus,
only the two equal-$k_{1}$ one-kink lightcone configurations
(each involving two of such coordinate patches) given in Fig. 2
are in principle possible.

Nevertheless, the fact that we have obtained just a unique, compact
expression for the time parameter $\tilde{t}$ should imply the
existence of just a unique one-kink lightcone configuration
involving all the four sets of
coordinate regions $k_{2}=k_{1}=+1$,
$k_{2}=k_{1}=-1$, $k_{2}=-k_{1}=+1$ and $k_{2}=-k_{1}=-1$.
Therefore, at least one of the poles at $r=0$ in one of the
configurations of Fig. 2 ought to be identified to the pole
having the same $\alpha$ and (in order to ensure the interval
for $\alpha$ to be ($0$,$\pi$))
different $k_{1}$ at $r=0$ in the other configuration. Inspection
of Eqns. (2.13), (2.15) and (3.8) tells us that, in fact, the
two poles in one of the two configurations should be identified
in this way to the corresponding two poles in the other
configuration. These identifications actually correspond by the
mapping $\theta\Rightarrow\alpha$ to the identifications made
in section 2 on the extreme values of real $\theta$. As it was
expected, there is thus a unique one-kink closed configuration
involving all four sets of
coordinate regions described by the physical
original regions $I^{k_{2}}_{k_{1}}$, $II^{k_{2}}_{k_{1}}$ in
Fig. 2. A rather pictorial representation of such a unique
lightcone configuration is given in Fig. 3. One can readily be
convinced of the existence of this configuration by proving
that if, along a null-geodesic itinerary, the above
identifications are made then, by starting at e.g.
$t=0$ on a given
surface $\phi=$const, $z$=const, $r=r_{0}$, $\tilde{t}=\tilde{t}_{0}$,
we finally recover the same point ($r_{0}$,$\tilde{t}_{0}$), also
at $t=0$, after completing the entire null-geodesic itinerary
along the lightcone configuration of Fig. 3.
We have checked by explicit calculation that this is
actually the case.
Insisting on the restriction of keeping $t=0$ along
a closed path requires equalizing points ($r$,$\tilde{t}$) on
pole identifications, while changing the sign of $\tilde{t}$ on
equator identifications.

\section{\bf Kruskal Extension of Extreme String Metric}
\setcounter{equation}{0}

The distinction between sign parameters $k_{1}$ and $k_{2}$
should manifest in the fact that $k_{1}$ must only be
involved in the definition of time $\tilde{t}$, and $k_{2}$
should appear in the definition of both time $\tilde{t}$
and radial coordinate $r$ as well. As already noted
before, that distinction could only be shown, even at $t=0$,
when the apparent singularity at $r=r_{*}$ is removed.

This geodesic incompleteness, which occurs in each of the
four sets of
coordinate regions described by metric (3.6), can be
removed by the usual Kruskal technique [21]. Thus, we
define the metric
\begin{equation}
ds^{2}=-2F(U,V)dUdV+dz^{2}+r^{2}d\phi^{2},
\end{equation}
in this way straightening the null geodesics into lines parallel
to the new $U$ and $V$ axis, and identify it with the standard
metric (3.6), with $g_{\tilde{t}\tilde{t}}=-\cos 2\alpha$,
$g_{\tilde{t}r}=-k_{1}k_{2}$ and $g_{UV}=-F$, in such a way
that $k_{1}$ and $k_{2}$ be kept consistently distinguished
and $F$ be finite, nonzero and depend on $r$ and $k_{2}$ alone.
All of these requirements can be met by the choice
\begin{equation}
U=\mp e^{\beta k_{1}\tilde{t}}\exp(2\beta k_{2}\int_{0/A}^{r}\frac{dr}{\cos
2\alpha})
\end{equation}
\begin{equation}
V=\mp\frac{1}{\beta r_{*}}e^{-\beta k_{1}\tilde{t}}
\end{equation}
\begin{equation}
F=\frac{r_{*}\cos 2\alpha}{2\beta}\exp(-2\beta
k_{2}\int_{0/A}^{r}\frac{dr}{\cos 2\alpha}),
\end{equation}
with $\beta$ a constant which should be chosen so that $F$ has a finite
limit as $r\rightarrow r_{*}$.

Using [19]
\[\int_{0/A}^{r}\frac{dr}{\cos
2\alpha}=\frac{1}{2}\ln(\frac{r_{*}+r}{r_{*}-r})\]
\[\cos 2\alpha=1-\frac{r^{2}}{r_{*}^{2}},\]
we obtain from (4.4)
\[F=\frac{(r_{*}^{2}-r^{2})}{2\beta
r_{*}}[\frac{(r_{*}+r)^{2}}{r_{*}^{2}-r^{2}}]^{-\beta k_{2}r_{*}}.\]

To avoid $F$ being either 0 or $\infty$ at $r=r_{*}$, we then choose
$\beta=-\frac{1}{k_{2}r_{*}}$, and arrive therefore at
\begin{equation}
F=-\frac{1}{2}k_{2}(r_{*}+r)^{2}
\end{equation}
\begin{equation}
U=\mp e^{-\frac{k_{1}k_{2}\tilde{t}}{r_{*}}}(\frac{r_{*}-r}{r_{*}+r})
\end{equation}
\begin{equation}
V=\pm k_{2}e^{\frac{k_{1}k_{2}\tilde{t}}{r_{*}}} ,
\end{equation}
so that
\begin{equation}
UV=-k_{2}\frac{(r_{*}-r)}{(r_{*}+r)}.
\end{equation}

In terms of the coordinate product (4.8) we have finally
\begin{equation}
F=\frac{2r_{*}^{2}k_{2}}{(k_{2}-UV)^{2}}
\end{equation}
\begin{equation}
r=r_{*}(\frac{k_{2}+UV}{k_{2}-UV})
\end{equation}
\[\tilde{t}=k_{1}t-k_{1}k_{2}r_{*}\{(1-\frac{4k_{2}UV}{(k_{2}-UV)^{2}})^{\frac{1}{2}}\]
\begin{equation}
+\frac{1}{2}\ln[\frac{(1+(1-\frac{4k_{2}UV}{(k_{2}-UV)^{2}})^{\frac{1}{2}})k_{2}UV}{1-(1-\frac{4k_{2}UV}{(k_{2}-UV)^{2}})^{\frac{1}{2}}}]\}.
\end{equation}
The metric (4.1) becomes then
\begin{equation}
ds^{2}=-\frac{4k_{2}r_{*}^{2}}{(k_{2}-UV)^{2}}dUdV+dz^{2}+\frac{r_{*}^{2}(k_{2}+UV)^{2}}{(k_{2}-UV)^{2}}d\phi^{2}.
\end{equation}

We notice that the $z$=const sections of this metric actually
coincide with that is obtained for a hemispherical section
($\theta=\frac{\pi}{2}$) of the De Sitter kink [19], the sole
though essential difference being that for each value of $k_{2}$
there are here two different values of time $\tilde{t}$
and $k_{1}t$,
corresponding to $k_{1}=\pm 1$, at every value of $r$ and $t$.
In Fig. 4 we give a representation in terms of coordinate $U$,
$V$ of the four different sets of
coordinate regions that occur in the
one-kink geodesically-complete extreme string spacetime.

The maximally-extended extreme string metric (4.12) does not
possess any singularity and covers all the space of a extreme
string, including both an unbroken-phase interior of radius
$r_{*}$ and a broken-phase covering shell of width
$(2^{\frac{1}{2}}-1)r_{*}$, which has been converted
into an unbroken phase containing a vanishing overall
real energy by the mapping $\theta\Rightarrow\alpha$
(see section 6).
The first of these facts is in
sharp contrast with the unavoidability of an unwellcome
singularity found by other authors for supermassive cosmic
strings [8,9]. The fact that (4.12) also describes an
exterior shell that protects the spherical core
from dissolving is, in turn,
in contrast with the trivial maximal extension (2.7) of the
interior metric of cosmic strings whose mass per unit length
is smaller than the critical value $G\mu_{c}=\frac{1}{4}$.
In the latter case, the topological defect need not developing
a broken-phase shell to protect itself against dissolving, as it
is always immersed in the broken-phase with conical geometry.

On the other hand, the finding of an extreme string kink
metric which on each $z$=const section exactly possesses the
symmetry of a hemispherical section of the De Sitter kink,
may have two consequences of interest. First, since the size
of the false vacuum region inside the extreme string
appears to exceed
the size of its corresponding horizon, the above symmetry
makes it consistent to consider the emergence of a unique
De Sitter inflationary process [22], without any fine tuning
of the initial conditions. Secondly, it will allow continuous
changes of topology from that of a two-sphere into that of
two hemispheres joined at their poles, and vice versa [23].
This may ultimately lead to the possibility for closed
timelike curves to occur throughout the maximally-extended
spacetime of an extreme string. Moreover, if we interpret
spherical extreme strings as incipient baby universes, then
the first of these implications leads, in turn, to quite
confortably accommodate the concept of an eternal process
of continually self-reproducing inflating universes [4,13]
into the present picture. The resulting model would, furthermore,
be implemented by the existence of a unique "genetic"
hallmark which can be expressed by the condition
$2G\mu=2G\eta^{2}=1$. For any other value of $\mu$ the
topological defect would not succeed in producing an
observable universe.

\section{\bf Closed Timelike Curves in an Extreme String}
\setcounter{equation}{0}

Closed timelike curves (CTCs) in general relativity were
already discussed by G"del, back in 1949 [24].
Such curves appeared in solutions which were shown to
require unphysical stress tensors [25]. The subject has
recently been revived in two different contexts. The first
involved tunnelling Lorentzian wormholes [26], and possesses
two essential drawbacks:
the used wormholes correspond to no vacuum
solution of Einstein equations, and moreover, they require
violation of the weak energy condition [26,27]. The second
recent attempt involves solutions to Einstein equations for
two rapidly moving, infinite parallel cosmic strings [28],
and has also met serious difficulties. In this case, the
CTCs can only appear at spatial infinity and, more importantly,
correspond to spacetimes with imaginary total mass [29].
In what follows of the present section and in section 6,
we shall discuss the possible existence of CTCs in the
spacetime of a single extreme string and address the
problem of their quantum characteristics.

It has already been mentioned that a single extreme string
kink may accommodate CTCs. In fact, inspection of Fig. 3
might already suggest that null geodesics could be CTCs
that somehow loop back through the new regions (i.e. regions
$III^{k_{2}}_{k_{1}}$ and $IV^{k_{2}}_{k_{1}}$ on
Fig.4), created by the extension process
with coordinates $U$, $V$, in addition to the original regions
(i.e. regions $I^{k_{2}}_{k_{1}}$ and $II^{k_{2}}_{k_{1}}$).
However, before showing in more detail how the CTCs arise in
the maximally-extended spacetime of a extreme string kink,
one should extend the surface identifications discussed
for metric (3.6) into the Kruskal language.

Clearly, to the conditions imposed for metric (3.6) (i.e. that
the values of $r$ and $\tilde{t}$ have to be the same on each
pair of identified points, and that whereas the value of
parameter $k_{1}$ should be preserved on equator identifications
and change on pole identifications, the value of parameter
$k_{2}$ must change on all identifications) we must now add the
two new conditions:
(1) Surfaces on original regions can only be
identified to surfaces on original regions, and (2) exactly the
same conditions that are applied for identifying surfaces on
original regions should also be applied to identify surfaces
on the new regions which are created by the Kruskal extension
(see Ref. [18]). The application of these conditions leaves
us with the following identifications on the extreme-string
Kruskal metric:

\noindent On the original regions of Fig. 4

$\bullet$ Upper hyperbola ($K_{2}=k_{1}=+1$)$\equiv$ Lower hyperbola
($k_{2}=-k_{1}=-1$)

$\bullet$ Upper hyperbola ($K_{2}=-k_{1}=+1$)$\equiv$ Lower hyperbola
($k_{2}=k_{1}=-1$)

$\bullet$ Right hyperbola ($K_{2}=k_{1}=+1$)$\equiv$ Right hyperbola
($k_{2}=k_{1}=-1$)

$\bullet$ Right hyperbola ($K_{2}=-k_{1}=+1$)$\equiv$ Right hyperbola
($k_{2}=-k_{1}=-1$)

\noindent On the new regions of Fig. 4

$\bullet$ Upper hyperbola ($K_{2}=-k_{1}=-1$)$\equiv$ Lower hyperbola
($k_{2}=k_{1}=+1$)

$\bullet$ Upper hyperbola ($K_{2}=k_{1}=-1$)$\equiv$ Lower hyperbola
($k_{2}=-k_{1}=+1$)

$\bullet$ Left hyperbola ($K_{2}=k_{1}=+1$)$\equiv$ Left hyperbola
($k_{2}=k_{1}=-1$)

$\bullet$ Left hyperbola ($K_{2}=-k_{1}=+1$)$\equiv$ Left hyperbola
($k_{2}=-k_{1}=-1$)

All other identifications being forbidden if we want to have one kink,
with $\alpha$ continuously varying along the entire interval
$0\leq\alpha\leq\pi$, in both directions. In particular, the
above identifications forbid the occurrence of any CTCs that
follow curved itineraries which are enforced to lie on the
$r$ axis at $t=0$, both in the original and new regions (see
Fig. 4).

The addition of the new regions, created in the maximal, nonsingular
extension [21] of the one-kink extreme string, leads to the
occurrence of two distinct one-kink lightcone configurations which
respectively associate with two different CTCs for geodesics which
start at time $t=0$, evolve along nonzero values of $t$ through
both, original and new regions, to finally loop unavoidably back
to $t=0$ again, at the same original value of $r$. These
evolutions have been checked by explicit calculation of the
physical time $t$ in all the situations where either $r=0$ or
$r=2^{\frac{1}{2}}r_{*}$ by taking into account the fact that
geodesic paths parallel to the $U$ axis preserve the value of
the time parameter $\tilde{t}$ unchanged, but the value of
$\tilde{t}$ will evolve according to
\begin{equation}
\tilde{t}(x_{i})=\tilde{t}(x_{j})+k_{1}k_{2}r_{*}\ln[\frac{(r_{*}+r(x_{j}))(r_{*}-r(x_{i}))}{(r_{*}+r(x_{i}))(r_{*}-r(x_{j}))}],
\end{equation}
with $x_{i}$ and $x_{j}$ two points on the trajectory, along geodesic
paths which are parallel to the $V$ axis. The value of the physical
time $t$ at each chosen value of $r$ ($r=0$ and $r=2^{\frac{1}{2}}r_{*}$)
was finally evaluated by the relation
\begin{equation}
t=k_{1}(\tilde{t}-\tilde{t}_{0}),
\end{equation}
where $\tilde{t}_{0}$ is the value of $\tilde{t}$ at $t=0$ calculated
by using (4.11) (or (3.10)).

The straight paths of the two null geodesics are represented on
the coordinate regions given in Fig. 4 by a solid line for the
CTC that starts at $r=0$ on the original region $I^{+}_{+}$
(a point we denote as {\it North}$_{+}$), and by a broken line
for the CTC that starts at $r=0$ on the original region $I^{+}_{-}$
(a point we denote as {\it North}$_{-}$). These geodesic itineraries
are also pictorially illustrated on Fig. 5, where we show how along
the CTCs a spacetime geometry which initially is e.g. typically
spherical evolves forth and back continuously into the geometry
of two hemispheres joined at their poles, to finally loop back
to the spherical geometry again. Also shown is Fig. 5 is the
evolution of the directions of lightcones on the hypersurfaces.
Finally, the itineraries and evolution of lightcone orientations
are represented on a $r$-$t$ diagram for the two CTCs in Fig. 6.

However, each of the CTCs crosses twice horizons at $V=0$ (i.e.
the $U$ axes in Fig. 4). At the crossing points, both $\tilde{t}$
and the physical time $t$ become either $\pm\infty$. It could
seem therefore that these sectors of the curves appear at time
infinity, and hence our CTCs become suspect of meetting
difficulties analogous to the spatial infinity
of the Gott's double-string
device [28]. Nevertheless, the fact that the energy of extreme
string becomes in the present approach quantized to a value
given by $(2G)^{-\frac{1}{2}}$ and, at the same time, the
two-sphere radius for metric (4.12) is $R=2^{\frac{1}{2}}r_{*}$
and the volume of the cylinder circunscribed to the two-sphere
is given by
\[V_{c}=4\pi R^{3}G\mu=2\pi R^{3},\]
implies that $r_{*}=G^{\frac{1}{2}}$. Then, since no string
with energy larger than $(2G)^{-\frac{1}{2}}$ may exist and real
energy has been assumed to be uniformly distributed
throughout the string interior, it follows that no localized
spacetime region with size smaller than $(2G)^{\frac{1}{2}}$
can be resolved inside the two-sphere. Therefore, only pairs
of points such as $r=0$ and $r=2^{\frac{1}{2}}r_{*}$,
separated a distance $(2G)^{\frac{1}{2}}$, can physically be
considered simultaneously along the CTCs.

Once points at $r=0$ and $r=2^{\frac{1}{2}}r_{*}$ have been
fixed on their respective hyperbolae, there will be no way
to obtain any physical information about any particular
trajectory joining such points. Thus, the ascending and
descending straight solid lines of Fig. 6 do not represent
real physical paths, but only directions of the
{\it quantum jumps} between the fixed points on the extreme
hyperbolae. These quantum transitions are only expressible
as propagators given by integrals over all possible
paths joining each pair of points on the hyperbolae. In a
small neighbourhood of $V=0$, these paths would correspond
to classical trajectories defined in the limit
$r_{*}\rightarrow 0$, where for real $U$ and $V$ we have
\[\mid U\mid=\lim_{r_{*}\rightarrow 0}e^{-k_{2}\frac{t}{r_{*}}}\]
\[\mid V\mid=\lim_{r_{*}\rightarrow 0}e^{k_{2}\frac{t}{r_{*}}}\]
so that the paths may cross the horizons $V=0$ at any finite values
of the physical time $t$.

The essential quantum nature of the spacetime of the extreme
string makes thus possible the existence of quantum transitions
between points of extreme hyperbolae on the original regions
and points of extreme hyperbolae on the new Kruskal regions,
and hence allow the occurrence of CTCs. These regions are
always mutually separated by a $V=0$ horizon on which quantum
theory allows the physical time to still keep any finite value.

According to the above results it appears that CTCs associated
with extreme strings are possible and do not show any of the
classical drawbacks that are present in the models mentioned at
the beginning of this section. The viability and causality
implications of these curves in a more complete quantum
treatment will be discussed in the next section.

\section{\bf The Quantum Extreme String}
\setcounter{equation}{0}

The procedure followed in section 2 in order to make topological
defects in gauge theories compatible with
gravitational topological defects (kinks) actually led to
quantization of the energy of supermassive cosmic strings.
Basically, this quantization arose from the specific
identifications of surfaces on the generally complex
interior spherical coordinate $\theta$ that result
from introducing one spacetime kink.
Two such identifications
were made. Identification of the surfaces corresponding to the
two extreme values of real $\theta$ implied that only the
energies associated with the critical and extreme strings
were allowed along the entire range of supermassive strings.
Identifications of the surfaces with maximum and minimum
values of imaginary $\theta$ at real $\theta=\frac{\pi}{2}$
did not provide however the string with any observable, real
energy, but only with a quantized exterior (true vacuum)
shell with the tachyonic energy.

It would be expected that once mapped onto the kink variable
$\alpha$, where even the evolution on imaginary $\theta$
becomes evolution on real $\alpha$ (Fig. 1), the true-vacuum
tachyonic energy be converted into real energy. Because of
the symmetry $\eta\rightarrow -\eta$ of the original broken
phase, there will be equal contributions to this energy, $E$, from
its positive and negative components.
However, although the overall energy of the string,
$(2G)^{-\frac{1}{2}}$, is thus
always conserved, the energy emerging from the symmetry
restoration in the mapping $\theta\Rightarrow\alpha$ can
still be stored in the different new
"unphysical" (inobservable) regions [30],
created in the distinct coordinate patches of the
Kruskal maximal extension, in such a way that if all of
its positive component goes to one of
such new regions for $k_{2}k_{1}=\pm 1$,
then all of its negative component goes to the same region for
$k_{2}k_{1}=\mp 1$, or to a different new region for $k_{2}k_{1}=\pm1$.
On the coordinate patches of Fig 4,
these would be, respectively, either regions $III^{\pm}_{\pm}$
and $IV^{\pm}_{\pm}$, or regions $IV^{\pm}_{\mp}$ and
$III^{\pm}_{\mp}$. For our present purposes, it will
suffice dealing with the storage of this energy in
regions $III$ only. The treatment for regions $IV$ is
completely parallel.

It appears that each of these energy components
could only emerge as an observable quantity in the
corresponding original exterior or interior region,
$II^{\pm}_{\pm}$ or $II^{\pm}_{\mp}$, if some hypersurfaces
in $III^{\pm}_{\pm}$ and $III^{\pm}_{\mp}$
are consistently identified with some
hypersurface in $II^{\pm}_{\pm}$ and $II^{\pm}_{\mp}$, respectively.
By a {\it consistent identification}
we mean here that identification which is predicted by the
geometry itself, not one which is introduced artificially
({\it ad hoc}), or by distorting the physical Lorentzian
time $t$ to the Euclidean imaginary region [31]. On the other
hand, since regions $III^{k_{2}}_{k_{1}}$
are essentially inobservable [30],
any manifestation of this energy distribution
in regions $II^{k_{2}}_{k_{1}}$ should
be in the form of a completely incoherent radiation which
brought no information whatsoever about regions $III^{k_{2}}_{k_{1}}$.
In what follows we shall show that such consistent identifications
actually exist for the coordinate regions given in Fig. 4. We will
also see that they give rise to a full protection against causality
violation along the complete paths of the CTCs discussed in section 5.

The identification required between surfaces in $III^{k_{2}}_{k_{1}}$
and $II^{k_{2}}_{k_{1}}$ (or between surfaces in $IV^{k_{2}}_{k_{1}}$
and $I^{k_{2}}_{k_{1}}$) will simply come from explicitely displaying
the sign ambiguity of the square root of the argument for the $\ln$
in Eqn. (3.10). Thus, if we want to express that argument as an
absolute value then one has as the most general expression for metrical
time
\[\tilde{t}(k_{3})\equiv\tilde{t}(k_{1},k_{2},k_{3})=\int_{0/A}^{r}d\tilde{t}=k_{1}t-r_{*}k_{1}k_{2}\{\frac{A}{r_{*}}(1-\frac{r^{2}}{2r_{*}^{2}})^{\frac{1}{2}}\]
\[-\ln\{\mid[\frac{(A(1-\frac{r^{2}}{2r_{*}^{2}})^{\frac{1}{2}}+r_{*})(r_{*}-r)}{(A(1-\frac{r^{2}}{2r_{*}^{2}})^{\frac{1}{2}}-r_{*})(r_{*}+r)}]^{\frac{1}{2}}\mid\}\}+\frac{i}{2}k_{1}k_{2}k_{3}(1-k_{3})\pi r_{*}\]
\begin{equation}
=\tilde{t}+\frac{i}{2}k_{1}k_{2}k_{3}(1-k_{3})\pi r_{*},
\end{equation}
where $\tilde{t}\equiv\tilde{t}(k_{1},k_{2})$ is the same as in (3.10),
and $k_{3}=\pm 1$ is a new sign parameter which unfolds the
coordinate regions given in Fig. 4 into still two sets of four
patches. Time (6.1) is the most general expression for the time
$\tilde{t}$ entering the standard metric (3.6). One would again
recover metric (3.6) from metric (4.1) with the same requirements
as in section 4 using $\tilde{t}(k_{3})$ instead of $\tilde{t}$
if we re-define the Kruskal coordinates $U$, $V$ such that
\begin{equation}
U=\pm
k_{3}e^{-\frac{k_{1}k_{2}\tilde{t}_{c}}{r_{*}}}\frac{(r_{*}-r)}{(r_{*}+r)}
\end{equation}
\begin{equation}
V=\mp k_{2}k_{3}e^{\frac{k_{1}k_{2}\tilde{t}_{c}}{r_{*}}},
\end{equation}
where
\begin{equation}
\tilde{t}_{c}=\tilde{t}+ik_{1}k_{2}k_{3}\pi r_{*}.
\end{equation}
This choice leaves expressions (4.8), (4.9), (4.10) and, of course,
the Kruskal metric (4.12) real and unchanged.

For $k_{3}=-1$, (6.2) and (6.3) become the sign-reversed to (4.6)
and (4.7); i.e. the points $(\tilde{t}-ik_{1}k_{2}\pi r_{*},r,z,\phi)$
on the coordinate patches of Fig. 4 are the points on the new
region $III^{k_{2}}_{k_{1}}$, on the same figure, obtained by
reflecting in the origins of the respective $U$,$V$ planes,
while keeping metric (4.12) and the physical time $t$ real
and unchanged. This is precisely the identification we required
to make the broken-phase tachyonic energy (which was first
quantized by identification of the extreme imaginary values
of $\theta$, and then stored in regions $III^{k_{2}}_{k_{1}}$
by mapping $\theta\Rightarrow\alpha$ and Kruskal extension)
appear in the original regions $II^{k_{2}}_{k_{1}}$ as an
observable real quantity. That this energy should appear in
$II^{k_{2}}_{k_{1}}$ as a thermal bath of completely incohent
radiation is now easy to see if one invokes the general relation
between metric periodicity and gravitational temperature [31,32].

This approach has, however, always seemed rather mysterious as
it gives no explanation to the use of the Euclidean version
of spacetime [33]. In our model periodicity in the metric
results, nevertheless, from a perfectly justifiable requirement
for mathematical completeness, rather than a suggestive though
not very convincing procedure, and entails no unjustified
extension of the physical time $t$ into the imaginary axis.
Our approach offers therefore a physically reasonable
resolution of the paradox posed by the Euclidean gravity
method, and provides the concept of gravitational temperature
with a new physical background - the relation between event
horizons and spontaneous symmetry breaking, discussed in
this paper.

As it was concluded in section 5, the evolution of a field
along any null geodesics in Fig. 4 should be described
by a quantum propagator, rather than a classical path.
If the field has mass $m$, such
a propagator will be the same as the propagator $G(x',x)$
used by Gibbons and
Hawking [34], and satisfy therefore the Klein-Gordon equation
\begin{equation}
(\Box_{x}^{2}-m^{2})G(x',x)=-\delta(x,x').
\end{equation}
For metric (4.12), the propagator $G(x',x)$ becomes analytic [34]
on precisely the strip of width $\pi r_{*}$ predicted by (6.4),
here without any need to extend $t$ to the Euclidean regime.
Then, the amplitude for detection of a detector [34] sensitive
to particles of a certain energy $E$, in regions $II^{k_{2}}_{k_{1}}$,
would be proportional to
\begin{equation}
\Pi_{E}=\int_{-\infty}^{+\infty}d\tilde{t}_{c}e^{-iE\tilde{t}_{c}}G(0,\vec{R'};\tilde{t}_{c},\vec{R}),
\end{equation}
where $\vec{R'}$ and $\vec{R}$ denote respectively $(r',z',\phi ')$
and $(r,z,\phi )$. Note that the time parameter entering the
amplitude for detection should be $\tilde{t}_{c}$, rather than
$t$, as both the matter field and the detector must evolve in
the spacetime described by metrics (3.6) and (4.12). Since time
$\tilde{t}_{c}$ (but not $t$) already contains the imaginary term
which is exactly required for the thermal effect to appear, we have
not need to make the physical time $t$ complex.
{}From (6.4) and (6.6), we can generally write
\begin{equation}
\Pi_{E}=e^{k_{1}k_{2}k_{3}\pi
r_{*}E}\int_{-\infty}^{+\infty}d\tilde{t}e^{-iE\tilde{t}}G(0,\vec{R'};\tilde{t}+ik_{1}k_{2}k_{3}\pi r_{*},\vec{R}).
\end{equation}

Following Gibbons and Hawking [34], we now investigate the different
particle creation processes that can take place on the extreme
hyperbolae at $r=0$ and $r=2^{\frac{1}{2}}$, along the quantum
itineraries of the CTCs discussed in section 5. Let us first
consider the case $k_{3}=-1$ for the original regions
$II^{k_{2}}_{k_{1}}$ on the patches of Fig. 4. For $k_{1}=k_{2}=+1$,
if $x'$ is a fixed point on the hyperbola at $r=0$ of region
$I^{+}_{+}$, and $x$ is a point on the hyperbola at
$r=2^{\frac{1}{2}}r_{*}$ of region $II^{+}_{+}$, we obtain
from (6.7)
\begin{equation}
P_{a}^{II^{+}_{+}}(E)=e^{-2\pi r_{*}E}P_{e}^{II^{+}_{+}}(E),
\end{equation}
where $P_{a}^{II^{k_{2}}_{k_{1}}}(E)$ generically denotes the
probability for detector to absorb a particle with positive
energy $E$ from region $II^{k_{2}}_{k_{1}}$, and
$P_{e}^{II^{k_{2}}_{k_{1}}}(E)$ accounts for the similar
probability for detector to emit the same energy also to region
$II^{k_{2}}_{k_{1}}$.

An observer on the extreme hyperbola of the exterior original
region of patch $k_{2}=k_{1}=+1$ will then measure an isotropic
background of thermal positive-energy radiation at a temperature
\begin{equation}
T_{s}=\frac{1}{2\pi r_{*}}.
\end{equation}

If, in turn, $x'$ and $x$ are fixed points on the extreme
hyperbolae in regions $I^{-}_{+}$ and $II^{-}_{+}$, respectively,
we obtain then for an observer on the hyperbola $r=0$ in the
interior original region
\begin{equation}
P_{a}^{II^{-}_{+}}(-E)=e^{2\pi r_{*}E}P_{e}^{II^{-}_{+}}(-E).
\end{equation}
According to (6.10), there will appear an isotropic background
of thermal radiation which is formed by exactly the antiparticles
to the particles contained in the thermal bath detected in
region $II^{+}_{+}$, at the same temperature $T_{s}$ given by
(6.9). The same probability relation as (6.10) is also
obtained for the exterior physical region $II^{+}_{-}$. Therefore,
observers on the extreme hyperbola at $r=2^{\frac{1}{2}}r_{*}$
in that region will also detect a thermal bath of particles
with energy $E<0$, at temperature (6.9). Finally, for the
point $x$ on the extreme hyperbola of the interior original
region $II^{-}_{-}$, we derive an expression as (6.8), also
for $E>0$, and hence an interpretation for thermal properties
as for region $II^{+}_{+}$. Had we similarly identified
extreme hypersurfaces of regions $I$ and $IV$, and then
carried out the parallel treatment for propagators $G(x',x)$,
the energy $E$ from $\theta\Rightarrow\alpha$ symmetry restoration
stored at regions $IV$ would have emerged as incoherent
radiation with particles of negative energy in regions
$I^{+}_{+}$ and $I^{-}_{-}$, and with particles with positive
energy in regions $I^{-}_{+}$ and $I^{+}_{-}$. Thus, the
overall balance of the energy being created in the original
regions would be exactly zero, with no correlation
(information) whatsoever being exchanged between any of
the different regions involved in the process. We note
that, since $r_{*}=G^{\frac{1}{2}}$, the temperature
(6.9) becomes $(2\pi G^{\frac{1}{2}})^{-1}$, i.e. close
to the Planck scale. Therefore, the emitted radiation is
in all the cases formed by a single particle with nearly
$\pm$ the Planck energy, which is uncorrelated to the
region it comes from.

For $k_{3}=+1$ we obtain similar hypersurface identifications
as for $k_{3}=-1$. In this case, the identification comes
about in the situation resulting from simply
exchanging the mutual positions of the original regions
$I^{k_{2}}_{k_{1}}$ and $II^{k_{2}}_{k_{1}}$ for,
respectively, the new regions $III^{k_{2}}_{k_{1}}$ and
$IV^{k_{2}}_{k_{1}}$, on the coordinate patches given in
Fig. 4, while keeping the sign of coordinates $U$, $V$
unchanged with respect to those in (4.6) and (4.7); i.e.
the points $(\tilde{t}+ik_{1}k_{2}\pi r_{*},r,z,\phi)$ on
the so-modified regions are the points on the original
regions $II^{k_{2}}_{k_{1}}$, on the same patches, again
obtained by reflecting in the origins of the respective
$U$,$V$ planes, while keeping metric (4.12) and the
physical time $t$ real and unchanged.

In the so generated new set of coordinate patches one can
now distinguish a new couple of CTCs, respectively starting
at the poles $South^{*}_{\pm}$ of the new regions. These
additional CTCs would exactly follow the same itineraries
as the CTCs considered in section 5 (see Fig. 4), but now
looping back through original physical regions, to finally
end at their respective starting points on the new regions.

Expressions for the relation between probabilities of absorption
and emission for $k_{3}=+1$ are obtained by simply replacing
regions $II^{k_{2}}_{k_{1}}$ for regions $III^{k_{2}}_{k_{1}}$
(or regions $I^{k_{2}}_{k_{1}}$ for regions $IV^{k_{2}}_{k_{1}}$),
and energy $E$ for $-E$, in (6.8) and (6.10), keeping the
same radiation temperature (6.9) in all the cases. Thus,
we obtain that observers on the extreme hyperbolae will detect
an isotropic thermal bath of particles with energy: (i) $E<0$
in the exterior new region $III^{+}_{+}$ and in the interior
new region $III^{-}_{-}$, and (ii) $E>0$ in the
interior new region $III^{-}_{+}$ and in the exterior new
region $III^{+}_{-}$, with the particle energies being
sign-reversed to these for radiation in the corresponding new regions
$IV^{k_{2}}_{k_{1}}$, in all the cases at an equilibrium
temperature (6.9). Thus, the overall energy created in the
new regions also vanishes.

Combining all the above results it follows that there exists
a close connection between the CTCs and the thermal processes
induced by the presence of the event horizon of an extreme
string. Along the entire CTC itinerary starting at $North_{+}$,
all the possible observers on the extreme hyperbolae of extrior
and interior (original and new) regions would always detect the
CTC by the presence of a totally incoherent radiation with
positive energy at the Planck temperature. No information whatsoever
could therefore be transferred from the CTC to such observers, or
{\it vice versa}, so spacetime causality could never be violated.
The situation for the second of the CTCs considered in section 5
is essentially the same, the only difference being that the
CTC starting at $North_{-}$ is always manifested by a thermal
radiation with negative Planck energy. For CTCs starting at
new regions one basically achieves identical conclusions,
with the CTC starting at $South^{*}_{+}$ giving always rise
to a thermal bath with negative Planck energy, and the CTC
starting at $South^{*}_{-}$ giving always rise to a thermal
bath with positive Planck energy, both at equilibrium temperature
(6.9) as well.

If no inflationary expansion occurred, temperature (6.9) and
radiated particle energy would always reach a value close to
the Planck scale. One would however expect the inflationary
process discussed in section 4 to take unavoidably place as
soon as an extreme string is formed. In this case, CTCs would
be associated to thermal processes quite less energetic.

In summary, the treatment carried out in the present section
seems to indicate that there is no quantum obstruction
preventing the existence of CTCs in possible cosmological
models based on minimally inflating extreme strings, in this
sense contradicting the Hawking's chronology protection
conjecture [35]. What the present model really prevents is
the possibility that any physically realistic observer may
get any experimental direct proof about the existence of
such curves. Or, re-paraphrasing Stephen Hawking [35], there
could perfectly be hords of tourists visiting us from the
future, but neither they nor we could know anything about their
trip. For them it would be a touring which costs a lot and
rewards nothing.

\vspace{1.5cm}

\noindent {\bf Acknowledgement}

The author thanks S.W. Hawking for hospitality in
the Department of Applied Mathematics
and Theoretical Physics, University of Cambridge, UK, where part
of this work was done, and G.A.
Mena Marug\'an, of IMAFF, CSIC, Madrid, for very enlightening
discussions. This work has been supported by a CAICYT Research
Project N§ PB91-0052.

\pagebreak

\noindent\section*{References}
\begin{description}
\item [1] T.W.B. Kibble, J. Phys. A9, 1387 (1976); A. Vilenkin,
Phys. Rep. 121, 236 (1985).
\item [2] A. Vilenkin, in {300 Years of Gravitation}, eds S.W. Hawking
and W. Israel (Cambridge Univ. Press, Cambridge, 1987).
\item [3] A. Vilenkin, Phys. Rev. Lett. 46, 1169 (1981);
Phys. Rev. D24, 2082 (1981); T.W.B. Kibble and N. Turok,
Phys. Lett. 116B, 141 (1982).
\item [4] A.D. Linde and D.A. Linde, Phys. Rev. D50, 2456 (1994).
\item [5] A. Vilenkin, {\it Topological Inflation}, gr-qc/940240,
preprint (1994).
\item [6] A. Vilenkin and E.P.S. Shellard, {\it Cosmic Strings
and Other Topological Defects} (Cambridge Univ. Press, Cambridge, 1994).
\item [7] A. Vilenkin, Phys. Rev. D23, 852 (1981).
\item [8] P. Laguna and D. Garfinkle, Phys. Rev. D40, 1011 (1989).
\item [9] M.E. Ortiz, Phys. Rev. D43, 2521 (1991).
\item [10] W.A. Hiscock, Phys. Rev. D31, 3288 (1985).
\item [11] J.R. Gott, Astrophys. J. 288, 422 (1985).
\item [12] D. Finkelstein and C.W. Misner, Ann. Phys. (N.Y.) 6, 230 (1959).
\item [13] A.D. Linde, $Inflation$ $and$ $Quantum$ $Cosmology$ (Academic
Press, Boston, 1990).
\item [14] C.W. Misner and A. Taub, Soviet Phys. JETP 28, 122 (1969).
\item [15] S. Deser and A.R. Steif, in {\it Directions in General
Relativity}, eds. B.L. Hu, M.P. Ryan Jr., and C.V. Vishveshwara
(Cambridge Univ. Press, Cambridge, 1993), Vol. 1.
\item [16] D. Kramer, H. Stephani, M. MacCallum and E. Herlt,
{\it Exact Solutions of Einstein's Field Equations} (Cambridge
Univ. Press, Cambridge, 1980).
\item [17] H.B. Nielsen and P. Olesen, Nucl. Phys. B61, 45 (1973).
\item [18] D. Finkelstein and G. McCollum, J. Math. Phys. 16, 2250 (1975).
\item [19] K.A. Dunn, T.A. Harriott and J.G. Williams, J. Math. Phys.
35, 4145 (1994).
\item [20] Note that the first term of the rhs in Eqn. (7) of
Ref. [19] contains the sign ambiguity here denoted as $k_{2}$.
This is incorrect and possibly due to a typing error.
\item [21] M.D. Kruskal, Phys. Rev. 119, 1743 (1981).
\item [22] A.H. Guth, Phys. Rev. D23, 347 (1981).
\item [23] The connection between kinks and topology change has
already been suggested by G.W. Gibbons and S.W. Hawking,
Phys. Rev. Lett. 69, 1719 (1992). Criticisms to this idea
were however raised by A. Chamblin and R. Penrose, Twistor
Newsletter 34, 13 (1992).
\item [24] K. G"del, Rev. Mod. Phys. 21, 447 (1949).
\item [25] V.P. Frolov and I.D. Novikov, Phys. Rev. D42, 1057 (1990).
\item [26] M.S. Morris, K.S. Thorne and U. Yurtsever, Phys. Rev.
Lett. 61, 1446 (1988); M.S. Morris and K.S. Thorne, Am. J. Phys.
56, 395 (1988).
\item [27] S.W. Hawking and G.F.R. Ellis, {\it The Large Scale
Structure of Space-Time} (Cambridge Univ. Press, Cambridge, 1973).
\item [28] J.R. Gott, Phys. Rev. Lett. 66, 1126 (1991).
\item [29] S. Deser, R. Jackiw and G. 't Hooft, Phys. Rev.
Lett. 68, 267 (1992).
\item [30] R.M. Wald, $General$ $Relativity$ (The University
of Chicago Press, Chicago, 1984).
\item [31] J.B. Hartle and S.W. Hawking, Phys. Rev. D31, 2188 (1976).
\item [32] S.W. Hawking, in {\it General Relativity. An Einstein
Centenary Survey} (Cambridge Univ. Press, Cambridge, 1979).
\item [33] N. Pauchapakesan, in {\it Hightlights in Gravitation
and Cosmology}, eds B.R. Iyer, A. Kembhavi, J.V. Narlikar and
C.V. Vishveshwara (Cambridge Univ. Press, Cambridge, 1988).
\item [34] G.W. Gibbons and S.W. Hawking, Phys. Rev. D15, 2738 (1977).
\item [35] S.W. hawking, Phys. Rev. D46, 603 (1992).

\end{description}

\pagebreak

\noindent {\bf Legends for Figures}.

\vspace{1cm}

\noindent $\bullet$ Fig. 1. Identifications ($\circ$) on
the orientable, complex spherical structure of an extreme
string, and mapping from that structure into that of the
corresponding one-kink. These identifications and mapping
lead to quantization of supermassive cosmic strings.

\vspace{.5cm}

\noindent $\bullet$ Fig. 2. Separate one-kink lightcone configurations
for the extreme string. Also represented are some geodesics passing
through the different regions $I^{k_{2}}_{k_{1}}$ and
$II^{k_{2}}_{k_{1}}$.

\vspace{.5cm}

\noindent $\bullet$ Fig. 3. Pictorial representation of
the one-kink lightcone configuration
connecting all the four possible spacetime regions $I^{k_{2}}_{k_{1}}$
and $II^{k_{2}}_{k_{1}}$ of the extreme string.

\vspace{.5cm}

\noindent $\bullet$ Fig. 4. The different coordinate regions of the
one-kink extended extreme string metric. In the figure,
$A=2^{\frac{1}{2}}r_{*}$, $\tilde{t}_{\alpha}=-r_{*}(2^{\frac{1}{2}}
+\ln(2^{\frac{1}{2}}-1))$, and $\tilde{t}_{\beta}=
r_{*}\ln(2^{\frac{1}{2}}-1)$. Each point on the diagrams represents
an infinite cylinder. The null geodesics starting at $North_{+}$
(solid line) and at $North_{-}$ (broken line) are two CTCs. The
solid curved lines at $t=0$ do not correspond to CTCs.

\vspace{.5cm}

\noindent $\bullet$ Fig. 5. Directions of the lightcones along
the CTC itineraries starting at (a) $North_{+}$ and (b)
$North_{-}$, showing the continuous change of spherical
symmetry into the symmetry of two hemispheres joined at their
poles.

\vspace{.5cm}

\noindent $\bullet$ Fig. 6. $r-t$ diagrams for the two CTCs,
starting at (a) $North_{+}$ and (b) $North_{-}$. The
directions of the lightcones given in Fig. 4 are also
represented. The transitions between hyperbolae on $r=0$ and
$r=2^{\frac{1}{2}}r_{*}$, diagramatically shown in the
figure as straight lines, actually represent the set of
all admissible quantum paths between $x$ and $x'$, at
either of such hyperbolae, for the path integral that
gives the propagator $G(x,x')$ [31,34] (see section 6).

\end{document}